# Incorporating the influence of sub-grid heterogeneity in regional-scale contaminant transport models

Boris Baeumer[1]

Yong Zhang[2]

Rina Schumer[3]

1. Department of Mathematics, University of Otago, Dunedin, New Zealand

2. Division of Hydrologic Sciences, Desert Research Institute, Las Vegas, NV 89119

3. Division of Hydrologic Sciences, Desert Research Institute, Reno, NV 89512

July 2013

**Abstract**: Numerical transport models based on the advection-dispersion equation (ADE) are built on the assumption that sub-grid cell transport is Fickian such that dispersive spreading around the average velocity is symmetric and without significant tailing on the front edge of a solute plume. However, anomalous diffusion in the form of super-diffusion due to preferential pathways in an aquifer has been observed in field data, challenging the assumption of Fickian dispersion at the local scale. This study develops a fully Lagrangian method to simulate sub-grid super-diffusion in a multi-dimensional regional-scale transport. The underlying concept is based on previous observations that solutions to space-fractional ADEs, which can describe super-diffusive dispersion, can be obtained by transforming solutions of classical ADEs. The

20  transformations are equivalent to randomizing particle travel time or relative velocity for each

21  model time step.  Here, the time randomizing procedure known as subordination is applied to

22  flow field output from MODFLOW simulations.  Numerical tests check the applicability of the

23  novel method in mapping regional-scale super-diffusive transport conditioned on local properties

24  of multi-dimensional heterogeneous media.

25  **Key Words:** Super-diffusion, Subordination, Regional flow, Lagrangian solver

26  **1. Introduction**

27      Groundwater contamination is a world-wide problem that frequently involves regional-

28  scale transport processes. The groundwater flow and transport models used to simulate and

29  predict contaminant transport are, by definition, simplifications of reality and can only account

30  for aquifer heterogeneity down to a finite level. Thus, the homogeneous grid cells of regional

31  flow models can be much larger than the Darcy, or theoretical representative elementary volume,

32  scale. Thus, local dispersion in a regional-scale transport model is typically assumed to be

33  Fickian, unless additional treatment (such as the kernel used in the convolution of solute

34  concentration [*Neuman and Tartakovsky*, 2009]) is used.

35      If we had perfect knowledge of aquifer heterogeneity and could develop perfect

36  representations of this heterogeneity in numerical simulations, then an equation for particle

37  advection would be all that is required to simulate transport.  The variability in particle transport

38  velocities would be driven by changes in hydraulic conductivity.  An advection-dispersion

39  equation (ADE) model allows for analytical or numerical incorporation of particle spreading

40  around the advective velocities. There are bounds, however, on how extreme or skewed

41  dispersion can be under the Fickian (classical ADE) model.  There have been different methods

42  proposed to address proper model treatment of various types of heterogeneity in groundwater,

including the upscaling approach [*Bellin et al.*, 2004; *Pruess*, 2004; *Zhang et al.*, 2010; *Zyvoloski et al.*, 2008] and the statistics based, downscaling method [*Lu et al.*, 2003; *Lu et al.*, 2002; *Ramanathan et al.*, 2008; *Ramanathan et al.*, 2010; *Sun et al.*, 2008]. Similar upscaling approaches have also been used for other environmental transport areas such as surface hydrology [*Barrios and Frances*, 2012; *Hansen et al.*, 2007] and atmospheric sciences [*Cassiani et al.*, 2010; *Karamchandani et al.*, 2009].

This study focuses on a new and relatively simple way of incorporating fractional advection-dispersion equation (fADE) models of transport into regional numerical models in order to move application of anomalous transport theory beyond academic study into application. The space fADE models are generalizations of classical ADE models that can parsimoniously incorporate extreme variation in particle velocities by replacing explicit representation of heterogeneity with a fractional derivative [*Meerschaert et al.*, 1999a]. It was previously observed that, more generally, the fractional power of the advection-dispersion operator leads to super-diffusion [*Baeumer et al.*, 2001] and its analytical solution could be obtained by transforming the solution to the classical ADE model. This transformation, known as subordination, amounts to randomizing the velocity of each particle for each time step. However in that model super-diffusive spread is present in all directions. One way to obtain governing equations whose solutions are super-diffusive in the direction of flow but diffuse normally in all other directions is to treat the fractional power of the advection operator and the diffusion operator separately; for the precise mathematical framework see [*Baeumer et al.*, 2009]. For a more general CTRW formulation that is faithful to the local velocity field, see *de Anna et al.* [2013].

While previous work demonstrated the subordination technique for analytical models, we adapt it here for use with numerical particle tracking techniques amenable to simulation of

66  transport subject to complex boundary and initial conditions. The goal is to apply subordination
67  techniques to the velocity field output from a MODFLOW [*Harbaugh et al.*, 2000] simulation
68  (or any other solutions of velocity using the finite difference method), thus incorporating sub-
69  grid heterogeneity into a particle tracking transport model. The method is coded into "Sub-
70  RWHet", which is a Lagrangian solver combining Darcy-scale non-Fickian diffusion with
71  coarsely resolved velocity fields. This particle tracking code is especially useful for evaluating
72  time to breakthrough at a sensitive receptor in cases where the classical ADE (applied to a coarse
73  regional grid) underestimates the early arrivals of contaminants.

74  The subordination method has also been used to produce sub-diffusion [*Baeumer and
75  Meerschaert*, 2001; *Schumer et al.*, 2003; *Sokolov and Metzler*, 2004] by changing the form of
76  the transformation. In this case, subordination gives a wide distribution of travel times,
77  producing a retardation-like effect with power-law tailing of breakthrough curves and works
78  independently to the spatial dispersion operator used in the physical model. This article focuses
79  on super-diffusion along flow lines and leaves extensions that include retardation effects for a
80  future study.

81  **2. Methodology development: Subordination in the direction of flow**

82  **2.1. Previous physical models**

83  We first review briefly the limitations of previous models in capturing sub-grid anomalous
84  super-diffusion. In one dimension the spatial fractional derivative has been used to capture
85  super-diffusion due to fast motion of solutes along preferential flow paths [*Benson et al.*, 2000a;
86  *Chaves*, 1998]

$$\frac{\partial}{\partial t}u(t,x) = -v\frac{\partial u(t,x)}{\partial x} + D\frac{\partial^\alpha u(t,x)}{\partial x^\alpha} \ , \tag{1}$$

88   with the initial condition $u(0,x) = u_0(x)$ and scale index $1 < \alpha \leq 2$. Here $v$ is the mean flow

89   velocity, $D$ is the dispersion coefficient, and $u$ denotes the solute concentration. When $\alpha = 2$,

90   the above space fADE reduces to the classical 2nd-order ADE.

91   In the fractional calculus framework, a mixing measure (denoted as $M$) has been used to

92   generalize this concept to higher dimensions by collating all the different fractional directional

93   derivatives [*Meerschaert et al.*, 1999b]

94   $$D_M^\alpha f(\mathbf{x}) = \int_{|\boldsymbol{\theta}|=1} \frac{d^\alpha}{ds^\alpha} f(\mathbf{x} + s\boldsymbol{\theta}) M(d\boldsymbol{\theta}), \qquad (2)$$

95   where different order derivatives in different directions are captured by a full operator-stable

96   fractional derivative $D_M^\alpha$ [*Meerschaert et al.*, 2001]. The variable $s$ in equation (2) denotes the

97   parameter describing distance travelled into the $\theta$ direction. One major drawback to these models

98   is that the direction $\theta$ is fixed in the integrand of the operator, and therefore $D_M^\alpha f$ is a mixture

99   of values of $f$ along *straight* lines to infinity irrespective of the conductivity field. While this

100  method produces reasonable particle trajectories in a multidimensional fractured rock setting

101  where the fracture orientation remains constant in space [*Reeves et al.*, 2008; *Zhang et al.*, 2010],

102  in general this is not desirable. A spatially dependent spectral measure requires straight line

103  jumps, which is not physically realistic, so it was also suggested that a streamline projection

104  method be used. In this method, the local dispersion coefficient at a point $x$ would determine the

105  total dispersive jump size over a time period $dt$ (irrespective of the local speed downstream as the

106  jump size distribution is based on the dispersion coefficient at the initial particle position) while

107  the jump direction varies with the flow field [*Zhang et al.*, 2006].

## 2.2. Development of the new physical model

We propose to model the influence of sub-grid heterogeneity not through randomizing jump distances but through randomizing *individual relative velocities*; i.e., having faster and slower particles. In other words, if we assume a particle following the flow at a given velocity is at position $X(t)$ by time $t$, an $s$-times faster particle would be already at position $X(st)$. The proportion of fast and slow particles is given according to the one-dimensional stochastic process that best models sub-grid heterogeneity in case of uniform flow. We call $\tau = st$ the *operational time* and the density of the one-dimensional stochastic process the *subordinator* and the resulting two or three dimensional process the *subordinated flow*.

It is important to note that when fast pathways are present, sub-grid heterogeneity might have a much bigger impact in the direction of flow than in other directions, while trapping of particles hinders movement of particles to any direction. Therefore we separate the random dispersive displacement (by subordination regional flow) from molecular (lateral) diffusion. The hydrodynamic dispersion contains mechanical dispersion and molecular diffusion. A larger local flux is more likely to transport the solute particle further downstream along the streamline. In other words, the super-diffusive jumps are not just related to the magnitude and direction of local flux but to the properties of the material that particles have to jump through.

We assume that the subordinating process is infinitely divisible, i.e., transport during uniform flow conditions on a grid level consists of many independent identically distributed random jumps. This implies that the Fourier transform of the subordinator $g$ satisfies

$$\frac{\partial}{\partial t}\hat{g}(t,k) = \psi(ik)\hat{g}(t,k); \hat{g}(t,k) = 1 \qquad (3)$$

129  where $\psi$ is the log-characteristic function of the subordinating process. For example, for

130  classical normalized advection (normalized to have velocity one, implying operational time is

131  equal to regular time), $\psi_1(ik) = -ik$, for classical normalized advection and dispersion,

132  $\psi_2(ik) = -ik + D(ik)^2$, for normalized advection and fractional dispersion,

133  $\psi_\alpha(ik) = -ik + D(ik)^\alpha, 1 < \alpha \leq 2$ [*Benson et al.*, 2000b], and for normalized advection and

134  tempered fractional dispersion $\psi_{\alpha,\lambda}(ik) = -(1+\alpha D\lambda^{\alpha-1})ik + D(ik+\lambda)^\alpha - D\lambda^\alpha, 1 < \alpha \leq 2$

135  [*Baeumer and Meerschaert*, 2010; *Cartea and del-Castillo-Negrete*, 2007; *Koponen*, 1995]. For

136  the case of fractional advection $\psi(ik) = -(ik)^\alpha, 0 < \alpha \leq 1$, subordinating in any direction was

137  already investigated in *Baeumer et al.*[2001]. Note that except for the first and last case, the

138  support of $g$ is the whole real line which means that negative operational times have to be

139  admissible. As only flow without dispersion is time reversible (going upstream), this restricts the

140  applicability of these subordinators to the flow direction.

141  If transport in the flow field is given by

142  $\frac{\partial}{\partial t}u(t,\mathbf{x}) = -\nabla.(\mathbf{v}(\mathbf{x})u(t,\mathbf{x})) = -\nabla_{\mathbf{v}(x)}u(t,\mathbf{x}); u(0,\mathbf{x}) = u_0(\mathbf{x})$, then the subordinated flow is

143  governed by

144  $$\frac{\partial}{\partial t}u(t,\mathbf{x}) = \psi\left(\nabla_{\mathbf{v}(\mathbf{x})}\right)u(t,\mathbf{x}); u(0,\mathbf{x}) = u_0(\mathbf{x}), \qquad (4)$$

145  where $\psi\left(\nabla_{\mathbf{v}(\mathbf{x})}\right)$ is a differential operator defined through a functional calculus [*Baeumer et al.*,

146  2009]. For example, if $\psi_2$ is as above,

147  $$\psi_2(\nabla_\mathbf{v}) = -\nabla_\mathbf{v} + D(\nabla_\mathbf{v})^2.$$

148  By definition, the solution to (4) is given by subordination; i.e. if $u_1$ is the solution to

149 $$\frac{\partial}{\partial t} u_1(t, \mathbf{x}) = -\nabla_{\mathbf{v}(\mathbf{x})} u_1(t, \mathbf{x}); u_1(0, \mathbf{x}) = u_0(\mathbf{x}) \tag{5}$$

150   and $g_\psi$ is the subordinator; i.e. $g_\psi$ solves (3), then

151 $$u_\psi(t, \mathbf{x}) = \int g_\psi(t, \tau) u_1(\tau, \mathbf{x}) d\tau \tag{6}$$

152   solves (4).

153   Now that we know how to handle subordination in the flow direction, we can add lateral
154   diffusion to the governing equation and try to solve

155 $$\frac{\partial}{\partial t} u(t, \mathbf{x}) = \psi\left(\nabla_{\mathbf{v}(\mathbf{x})}\right) u(t, \mathbf{x}) + \nabla.D\nabla u(t, \mathbf{x}); u(0, \mathbf{x}) = u_0(\mathbf{x}). \tag{7}$$

156   Model (7) is the new physical model proposed by this study. In the following we approximate
157   and apply it.

158   The Markovian model governed by (7) can now be further extended to include particle
159   trapping, mobile/immobile extensions, reaction terms, etc., by using, for example, fractional
160   temporal derivatives. This is straight forward as the operator on the right hand side of (7) is still a
161   tractable linear generator of an analytic, dissipative semigroup. Since this study focuses on
162   super-diffusion, we leave properties of these extensions of (7) to a future study.

163

164   **2.3. Numerical approximations**

165   The solution of model (7) can be approximated using the operator splitting method. First,
166   approximate by time stepping, advancing in each time step according to (7) with $D=0$ and $u_0(\mathbf{x})$
167   being the current state, and then according to (7) with $\psi = 0$ [*Baeumer et al.*, 2009]. For faster

168 convergence, the order can be alternated [*Strang*, 1968]. Instead of subordinating to obtain the

169 solution to (7) with $D=0$ at a small time step, one can use a numerical approximation of

170 $\psi(\nabla_{\mathbf{v}(\mathbf{x})})$ and approximate (implicitly) via $u_\psi(dt,\mathbf{x}) = u_0(\mathbf{x}) + dt\psi(\nabla_{\mathbf{v}(\mathbf{x})})u_\psi(dt,\mathbf{x})$. Numerical

171 approximations of $\psi(\nabla_{\mathbf{v}(\mathbf{x})})$ can be obtained by transferring shift invariant numerical

172 approximations of $\psi\left(\dfrac{d}{dx}\right)$ onto flow lines; i.e. if $\psi\left(\dfrac{d}{dx}\right)f(x) = \lim_{h\to 0}\sum_j w_{h,j}f(x-y_{h,j})$ for

173 some weights $w_{h,j}$ and shifts $y_{h,j}$, then

$$\psi(\nabla_{\mathbf{v}(\mathbf{x})})(f)(\mathbf{x}) = \lim_{h\to 0}\sum_j w_{h,j}u_1(y_{h,j},\mathbf{x}) \qquad (8)$$

175 [*Baeumer et al.*, 2009]. The solution can also be approximated by using a particle tracking code

176 such as the one described below, alternatively jumping along the flow line for a time period *dt*

177 with relative particle speed given by the subordinator, and jumping for a time period *dt* according

178 to the multivariate Brownian motion with generator $\nabla\cdot D\nabla$.

### 3. Particle tracking approach

180 We propose a 4-step Lagrangian scheme to subordinate regional-scale flow, after

181 MODFLOW is used to solve the steady-state or transient flow field. We subordinate according to

182 the most general model mentioned in Section 2, the tempered stable model:

$$\psi(ik) = \psi_{\alpha,\lambda}(ik) = -(1+\alpha D\lambda^{\alpha-1})ik + D(ik+\lambda)^\alpha - D\lambda^\alpha, 1<\alpha\leq 2.$$

184 The following particle-tracking algorithms are coded into RWHet [*LaBolle*, 2006], which is a

185 mature Random Walk solver for simulating solute transport in heterogeneous porous media that

186 has undergone significant model validation [*Zhang et al.*, 2012]:

187    1) Calculate the operational time for each particle during the *k*-th jump:

$$\delta t_k = dt_k + \left(\sigma_R dt_k\right)^{1/\alpha} dL_{\alpha,\lambda}(\beta = +1, \sigma = 1, \mu = 0) \tag{9}$$

where $dt_k$ [T] denotes the user-defined time step (which is adjusted further in RWHet to account for the boundary condition and transient flow, etc.), $\sigma_R [T^{\alpha-1}]$ is a measure of uncertainty and $dL_{\alpha,\lambda}$ [dimensionless] is an $\alpha$–stable tempered random variable with skewness $\beta = +1$, scale $\sigma = 1$, shift $\mu = 0$, and truncation parameter $\lambda$ (see *Baeumer and Meerschaert* [2010] on generation techniques).

   2) Track the advective displacement of the particle along the present stream lines for the operational time $\delta t_k$ [T]. Note that we may need to divide $\delta t_k$ into sub-time steps (depending on the magnitude of local velocities) to remain on the flow line, and relocate the particle around the grid interface. In addition, the operational time $\delta t_k$ can be either positive or negative, since the scale index $\alpha$ is large than 1. Hence we reverse the flow field and track the backward motion of the particle when $\delta t_k < 0$.

   3) Calculate the lateral dispersion (i.e., the concentration-gradient driven diffusion) for each particle during the time step $dt_k$ using the Itô-Euler integration scheme [*Gardiner*, 1985]:

$$dX_i = \left[\frac{\partial}{\partial x_i} D_{ij}(X,t)\right] dt_k + B_{ij}(X,t) d\omega_j(dt_k) \tag{10}$$

where $i, j = 1,2,3$ denotes the three axes in space, $\omega_j [T^{1/2}]$ (a Wiener process) is a vector of independent normally-distributed random variables with zero mean and variance $\langle d\omega_i, d\omega_j \rangle = \delta_{ij} dt_k$, $B_{ij}$ [$LT^{1/2}$] is a tensor defining the strength of diffusion ($2D_{ij} = B_{ik} B_{kj}$).

206   Here we assume that the medium porosity is constant. This step has already been coded into
207   RWHet [*LaBolle*, 2006] by its author.

208   4) Update particle and grid properties, and iterate steps 1-3 until reaching the end of
209   simulation, or until all particles exit the model domain.

210   **4. Numerical experiments: Validation and application**

211   We validated extensively the methods described above. Two numerical experiments are
212   shown below.

213   **4.1. Example 1**

214   This example non-trivially tests the 1-*d* concentration profile (snapshot) simulated by Sub-
215   RWHet along a straight streamline with local average velocity, which is two orders of magnitude
216   lower around x=-1/2 and x=3/2 than at other locations due to, for example, a change in porosity
217   (see Figure 1a). The transport model (7) along the x-axis is now reduced to

218 $$\frac{\partial}{\partial t}u(t,x) = -V(x)\frac{\partial}{\partial x}u(t,x) + \sigma_R \left[V(x)\frac{\partial}{\partial x}\right]^\alpha u(t,x), \quad (11)$$

219   where we take $\sigma_R = 0.1$ for this example, and the lateral diffusion coefficient *D*=0 is ignored to
220   focus on the subordinated flow. The analytical solution on the infinite domain is given by

221 $$u(t,x) = \frac{1}{(\sigma t)^{1/\alpha}} g_\alpha\left(\frac{T_x - t}{(\sigma t)^{1/\alpha}}\right); \quad T_x = \int_{x_0}^{x} \frac{1}{V(\xi)} d\xi$$

222   where $g_\alpha$ is the standard skewed $\alpha$-stable density with Fourier transform $\hat{g}(k) = e^{(ik)^\alpha}$. The
223   analytical solution can be accurately approximated using standard codes for computing $\alpha$-stable
224   densities.

225   The numerical model consists of 400 columns, 1 row, and 1 layer. The cell width along
226   rows is $\Delta x$=0.01, the cell width along column is $\Delta y$=1, and the layer thickness is $\Delta z$ =1. An
227   instantaneous point source (with $10^6$ particles) at time $t$=0 is located at position $x_0 = -1.5$. The
228   1-$d$ model domain is [-2, 2]. For the transport model, the left boundary (as $x = -2$) is treated as a
229   reflective boundary, where the particle exiting the left boundary (due to backward transport,
230   since the flow is from left to right) is reflected back to the model domain. The right boundary is
231   an absorbing boundary. To account for the change in porosity the particle count needs to be
232   normalized by dividing it by the porosity. In a 2-$d$ divergence free (constant porosity) field this
233   would automatically be achieved by a varying flow line density. Numerical solutions of Sub-
234   RWHet generally match the analytical solutions (Fig. 1b,c). The noise in the simulated snapshot
235   with low concentrations can be improved by either increasing the number of particles, or using
236   the particle splitting technique. Note not only the presence of downstream concentration, where
237   there is none in the α=2 case, but also the drop-off in concentration after the slow zones
238   highlighting that slow zones affect the fractional dispersion; i.e. even particles with high relative
239   velocities are going slowly.

240

241   **4.2. Example 2**
242   Example 2 considers flow and transport in a 2-$d$ heterogeneous aquifer, to test the
243   applicability of Sub-RWhet to solve complicated problems where either the analytical solutions
244   or other numerical solvers are not available. The flow model configuration is shown in Fig. 2a,
245   where ground water flows from top to bottom (bounded by the two vertical no-flow boundaries).
246   Water is injected (at a rate $Q$=0.09 m$^3$/s) into the aquifer through a fully penetrating well near the
247   top boundary, and an extraction well (at a pumping rate $Q$=1.16 m$^3$/s) is located near the bottom

248  boundary. The two wells are separated by a low-permeability zone whose hydraulic conductivity
249  is 3-orders of magnitude smaller than the surrounding high-permeability material. A similar
250  model setup was used by *Zheng and Wang* [1999].

251  We first check whether the subordinated particle trajectories follow streamlines. Results
252  show that Sub-RWHet captures precisely the position of random-walking particles along
253  streamlines at various times (Fig. 2b~d). The model solved for this case is equation (7) without
254  the lateral dispersion term (the last term on the right-hand side of (7)). The calculated advective
255  displacement of each particle generally follows the streamlines, which diverge or converge due
256  to the injection and pumping of groundwater.

257  We then check whether Sub-RWHet can capture the super-diffusive transport. A
258  continuous source with unit concentration is injected at the injection well. Both RWHet and
259  Sub-RWHet are used to track contaminant particles through the heterogeneous medium. The
260  water injection rate for this case is changed to 0.001 $m^3$/s, and the extraction rate at the pumping
261  well is 0.0189 $m^3$/s (as in *Zheng and Wang* [1999]). The simulated particle plumes are shown in
262  Fig. 3, where the subordinated plume has a relatively faster leading edge, as expected.

263  **5. Conclusions**

264  In cases where ADE-based numerical models overestimate solute travel times for the
265  leading edge of a plume, alternative models that can simulate super-diffusion are needed for
266  more accurate risk characterization. This study derives a novel physical model (eq. (7)) which
267  combines local and nonlocal transport by subordinating regional flow. The sub-grid dispersion is
268  not limited to the Fickian case, but can rather be a wide range of dispersive processes varying
269  from super-diffusion to normal diffusion. Numerical approximations using both the operator-
270  split time-stepping method and the particle tracking method are provided, where numerical experiments of the latter are shown for demonstration. The resultant Lagrangian solver "Sub-RWHet" accounts for the sub-grid super-diffusion, and it can read grid-based velocity fields generated for example by the popular USGS code MODFLOW. An executable Sub-RWHet is available upon request from the authors.

It is noteworthy that the new physical model (7) and its solver extend the previous approach for modeling space-dependent anomalous super-diffusion [*Zhang et al.*, 2006]. The dispersion coefficient in model (7) is not limited to a constant, but can vary in space and/or time. To the best of our knowledge, it is therefore the first Lagrangian model that allows multi-dimensional super-diffusion conditioned on local aquifer properties. On the other hand, model (7) and the other fractional-derivative models are now limited to fitting models, where the quantitative linkage between medium heterogeneity and model parameters remains to be shown. We will focus on the model predictability in a future study.

**Acknowledgement**: We would like to thank Dr M. Kovács for many helpful discussions. YZ was supported by the Desert Research Institute and the National Science Foundation under Grant DMS-1025417. YZ thanks Dr. E. M. LaBolle for providing his code RWHet. The copyright of the original RWHet belongs to LaBolle. This paper does not necessarily reflect the views of the funding organizations.

399 Figure 1

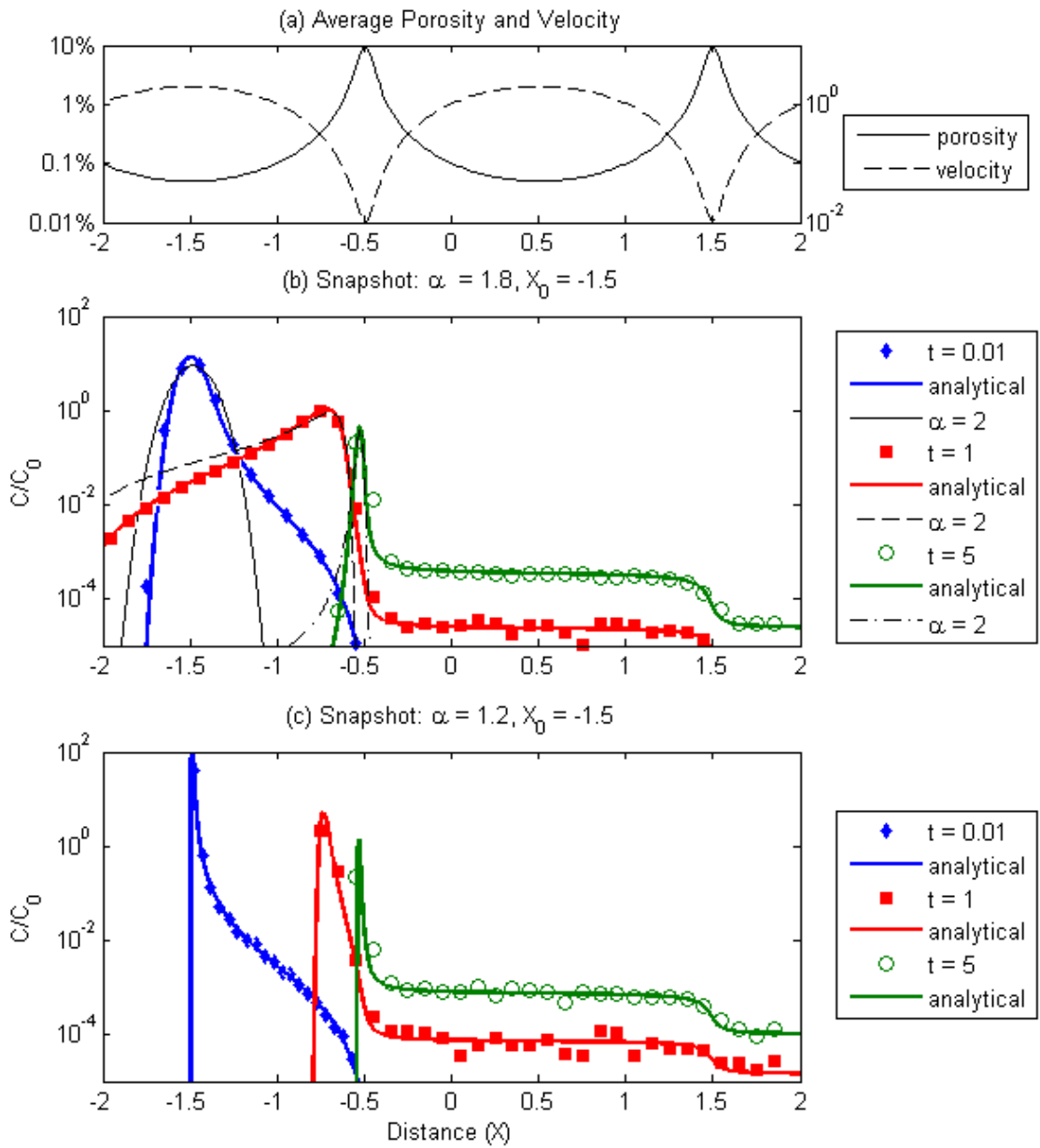

Figure 1. Example 1: Numerical solutions of Sub-RWHet (11) versus the analytical solution obtained by subordinating the solution to the advection equation for 1-d tracer snapshots for $\alpha=1.8$ (b) and $\alpha=1.2$ (c), respectively, for various times (t=0.01, 1, and 5) injected at $x=-1.5$. Notice the heavy-tailed down-stream concentration being influenced by local low velocity zones. (a) shows the porosity and velocity.

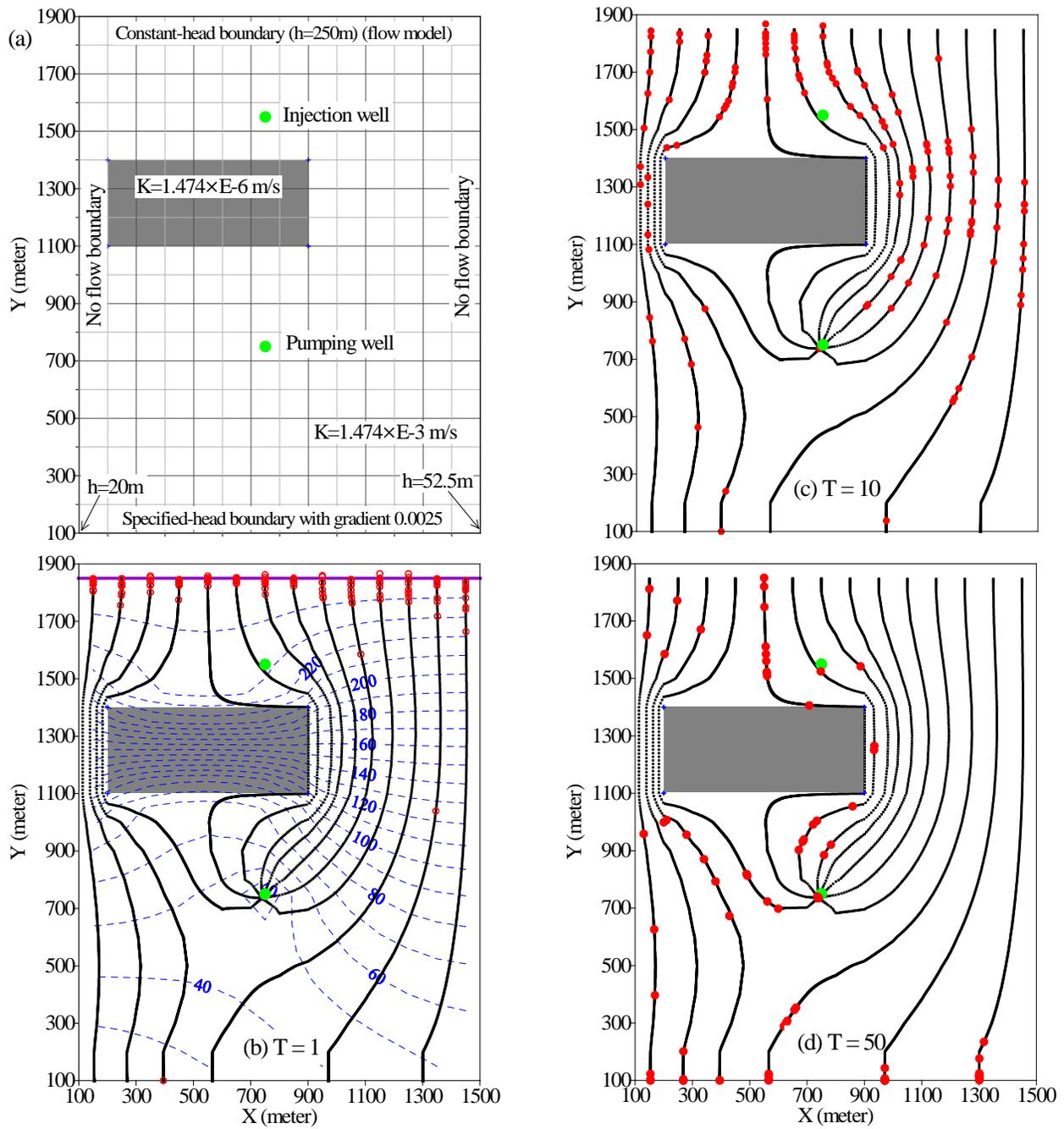

407

408 Figure 2. Example 2: Numerical tests of Sub-RWHet: position of particles (red dots) at time T=1

409 day (b), 10 days (c), and 50 days (d). (a) shows the 2-d flow model. In (b), the dashed lines

410 denote the simulated hydraulic head using MODFLOW. The solid lines in (b), (c) and (d)

411 are the streamlines calculated by RWHet. The green dots denote the location of two wells.

Figure 3

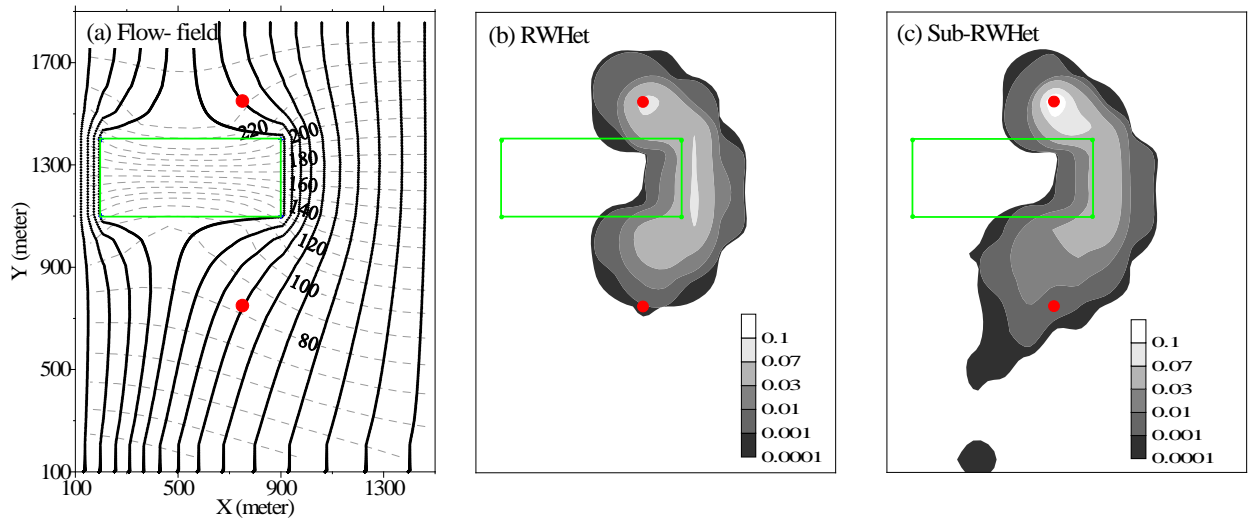

Figure 3. Example 2: The simulated particle plume (normalized concentration) at time T=10 days using RWHet (b) and Sub-RWHet (c), respectively, in a 2-d flow field (a).